# Quantifying the Impact of Upright Patient Positioning on Cardiac Substructures Using Deep Learning

## Authors:

Nicholas Summerfield, PhD[1,2]

Yuhao Yan, PhD[1,2]

Chase Ruff, MS[1,2]

Mark Pankuch, PhD[3]

Shae Gans[3]

Niek Schreuder, PhD[4]

Carri K Glide-Hurst, PhD[1,2]

**Affiliations:**

1. Department of Radiation Medicine, University of Wisconsin-Madison, Madison, Wisconsin
2. Department of Medical Physics, University of Wisconsin-Madison, Madison, Wisconsin
3. Northwestern Medicine Proton Center, Warrenville, Illinois
4. Leo Cancer Care, Inc., Middleton, Wisconsin, USA

**Corresponding Author:** Carri Glide-Hurst, glidehurst@humonc.wisc.edu, Dept of Radiation Medicine, 600 Highland Avenue, K4/B100-0600, Madison, WI 53792

**Statistical Author:** Nicholas Summerfield, nsummerfield@wisc.edu, Dept of Radiation Medicine, 600 Highland Avenue, K4/B100-0600, Madison, WI 53792

**Funding statement:** Work reported in this publication was supported in part by the National Cancer Institute of the National Institutes of Health under award numbers R01HL153720 (PI:

Carri Glide-Hurst). The content is solely the responsibility of the authors and does not necessarily represent the official views of the National Institutes of Health.

**Data sharing statement:** Data is not available at this time.

**Acknowledgements:** The authors would like to thank the Northwestern Medicine Proton Center for their help in acquiring patient datasets.

## Abstract

**Purpose:** Upright patient positioners with diagnostic-quality vertical CT at treatment isocenter may improve image-guided particle and radiation therapy (RT) by increasing lung volume, reducing tumor motion, and improving cardiac sparing. However, cardiac substructure (CS) geometry in upright patients remains insufficiently characterized. This work evaluated whether a supine-trained deep-learning (DL) CS segmentation model generalizes to upright CT images and quantified posture-dependent CS positional changes in thoracic patients, examining whether upright positioning produces meaningful changes that may support future cardiac-sparing workflows.

**Methods:** Eight thoracic proton therapy patients underwent paired supine and upright 4DCT imaging. Lung volumes were compared for posture-dependent changes. Twenty CS, including whole heart, chambers, great vessels, coronary arteries, valves, and conduction nodes, were manually labeled on both datasets. A previously developed supine-trained nnU-Net-based DL pipeline generated the same CS, and performance was evaluated using Dice similarity coefficient (DSC) and 95% Hausdorff distance (HD95). Upright CTs were rigidly registered to corresponding supine CTs by aligning the thoracic vertebrae, and CS centroid shifts were measured in the registered coordinate frame and relative to the carina. Paired differences were assessed using Wilcoxon signed-rank tests, with $p<0.05$ considered significant

**Results:**

The DL model successfully predicted all 20 CS on both upright (DSC, 0.65±0.24; HD95, 7.6±5.6mm) and supine (DSC, 0.72±0.19; HD95, 5.8±2.6mm) orientations yet with lower ($p<0.05$) performance on upright. Upright positioning significantly increased median lung volume by 20.6% (range, -8.9%-42.8%). After vertebral alignment, most CS centroids shifted significantly inferior (median WH shift, 23mm; range, 18-37mm) and anterior (median WH shift, 5.0mm; range, 1.0-13.0mm) when upright. Relative to the carina, most CS shifted significantly inferior and closer to in-line anterior-posterior alignment.

**Conclusions:** A supine-trained DL model generalized to upright CT images for CS segmentation. Upright positioning produced increased lung volumes and significant inferior CS displacement, suggesting favorable geometry changes that may support cardiac-sparing workflows.

## Introduction

Thoracic radiation therapy (RT) and particle therapy (PT) are challenged by respiratory and cardiac motion[1], which can affect target geometry, organ-at-risk positioning, and setup reproducibility across treatment fractions[2]. Data obtained from patients being treated at deep or moderate inspiration breath-hold demonstrates that increasing lung volumes can reduce respiratory motion and improve thoracic treatment geometry, normal tissue sparing, and inter-fraction repeatability[3,4]. Recently, novel upright-patient positioners coupled with diagnostic-quality CT have enabled PT and RT in an upright position and been shown to offer several potential advantages for patient treatment[5], including alternative treatment geometries and reduced respiratory motion[6], while emerging evidence also suggests that upright RT may improve patient comfort and emotional well-being[7]. Similarly, upright systems enable gantry-free setups, reduced construction requirements, and introduce higher-quality CT imaging at treatment isocenter in RT[5]. Yet, established modern RT protocols and technologies were developed primarily for supine patient positioning. Therefore, additional investigation is needed to determine how upright positioning affects treatment geometry and whether computational methods developed for supine workflows remain reliable when applied to upright imaging.

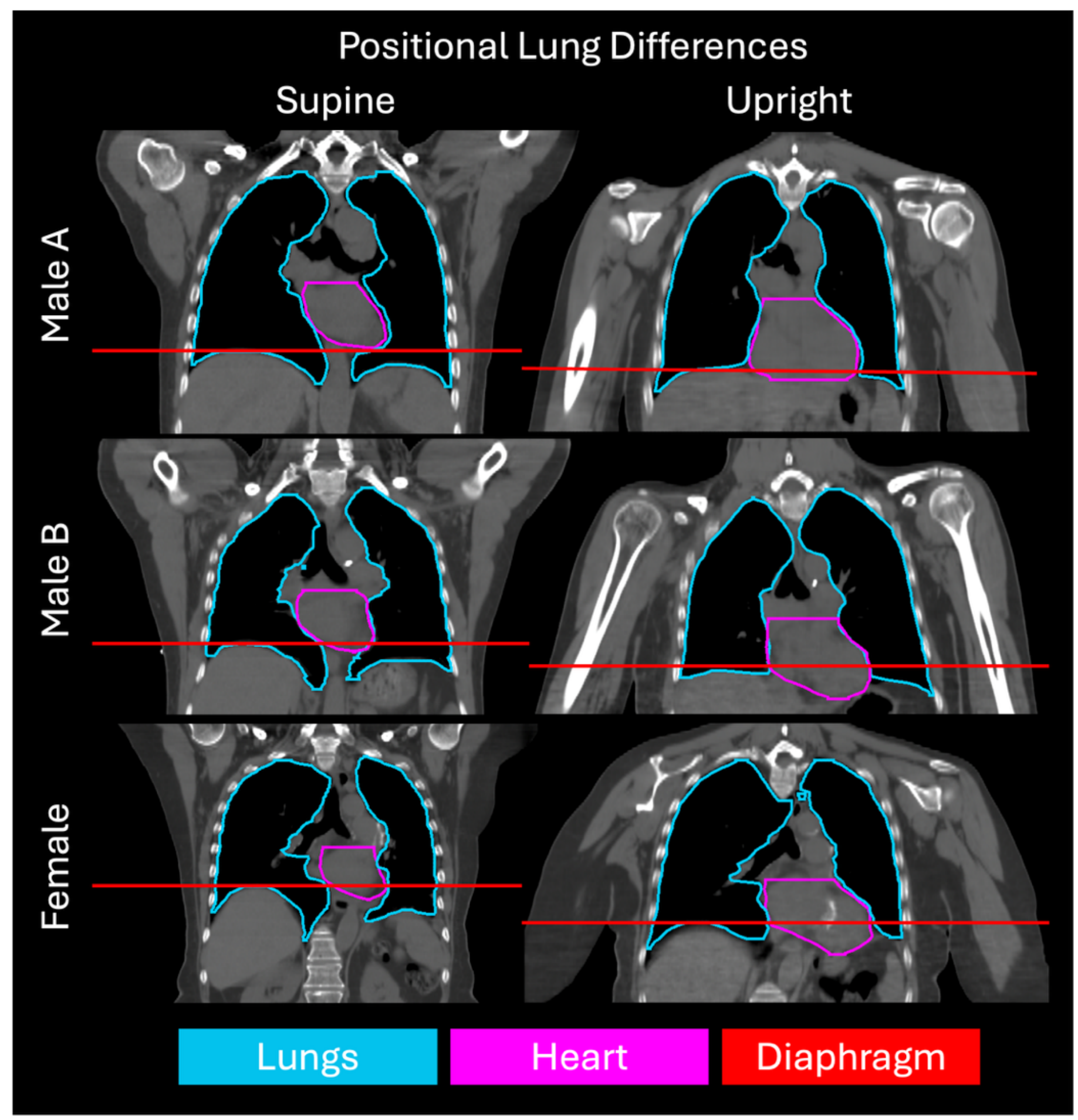

Figure 1: Representative paired supine and upright average 4D CT images for three arbitrary patients showing coronal slices, aligned with the superior lung extent, demonstrating increased lung volume towards the diaphragm when upright.

Upright positioning has previously been investigated and is associated with systematic changes in thoracic anatomy. Recent studies have reported larger lung volumes in the upright position[5,6,8,9], as shown in Figure 1, with observed reductions in intrathoracic respiratory motion[6], suggesting potential advances for thoracic treatment geometry and normal lung sparing. RT-focused paired upright and supine 4DCT analysis have similarly demonstrated increased lung volumes and measurable, primarily inferior displacement of the whole-heart when upright[8]. Furthermore, Martire *et al.*[10] demonstrated in a 4D carbon-ion planning comparison that upright positioning may create patient-specific planning opportunities, including improved target coverage, reduced upright motion, and improved heart sparing for select patients when beam geometries were adapted to the upright anatomy. These observations motivate further evaluation of whether the lung-volume and thoracic-geometry changes observed in the upright position translate into more favorable cardiac geometry for thoracic RT and PT.

Historically, the heart has been treated as a single organ-at-risk (OAR) during RT treatment planning, using simple whole-heart dose metrics, such as mean heart dose, to quantify radiation-induced heart disease. However, whole-heart metrics do not capture local dose deposition[11] and therefore do not adequately describe risk of individual cardiac substructures (CS) such as the coronary arteries, chambers and valves[12]. Moreover, the CS dosimetric endpoints are more strongly correlated with late-toxicities and morbidities post-treatment than the whole-heart alone[13]. Retrospective studies have associated dose to specific CS with outcomes such as coronary calcification, left ventricular dysfunction, and heart failure[13], while planning studies have demonstrated the feasibility of incorporating CS constraints into thoracic RT workflows for improved cardiac sparing[14,15].

Due to the large number of radiosensitive CS and the contouring burden associated with numerous small structures, many studies have investigated deep-learning (DL) auto-segmentation tools as a practical strategy for incorporating CS into RT workflows[16]. Yet, the images used to train DL models are commonly acquired from standard RT simulation scans or diagnostic scans[16] acquired in the supine position. Models trained primarily on supine images may experience domain shift when applied to upright images, where gravity produces systematic changes in organ position, soft-tissue deformation, and thoracic geometry such as increased lung volumes[17]. These systematic anatomical differences may create more favorable geometry for thoracic tumors near the heart, while also altering the CS position and shape, thereby potentially impacting the generalizability of supine-trained CS autosegmentation models.

Despite growing interest in upright RT, the effects of upright positioning on the heart and cardiac substructure position and on the performance of supine-trained CS auto-segmentation models remain insufficiently characterized. We present a quantitative analysis of paired supine and upright datasets of patients undergoing thoracic proton therapy (PT) to evaluate the performance of DL-based CS auto-segmentation across setup conditions and characterize position-dependent cardiac and CS displacement to lay the groundwork for potential advantages in cardiac sparing using upright RT.

## Methods

### Patient population

Eight patients (five male, three female) undergoing PT for advanced thoracic, central, or ultra-central targets who received both supine and upright computed tomography (CT) imaging were retrospectively evaluated via the Proton Collaborative Group (PCG) from a multi-institutional registry capturing real-world clinical and dosimetric information from patients treated with proton therapy. All patients underwent respiratory phase-sorted 4D CT in the supine and upright positions. Supine imaging was acquired with both arms elevated above the head using a LightSpeed RT16 (GE Medical Systems, Waukesha, WI) CT simulator (CT-SIM) (120kVp, 1.27mm in-plane resolution, 2.5mm slice thickness). Upright imaging was acquired with both arms relaxed inferiorly alongside the torso using an upright patient positioner (P-Cure Inc., Akron, OH) coupled with a wall-mounted Brilliance Big Bore (Philips Medical Systems, Cleveland, OH) CT-SIM (120kVp, 1.17mm in-plane resolution, 3mm slice thickness) at a 20° tilt angle. For each patient and position, the respiratory phase-resolved 4DCT was averaged across phases to generate an average CT image. These average CTs served as the primary reference images for evaluating posture-dependent differences in gross anatomy and cardiac substructure positions, consistent with the image representation used for proton treatment planning in these cases. For a subset of three patients, the end-inhalation and end-exhalation phase images were also available in both upright and supine positions and were used for an exploratory analysis of respiratory-induced CS displacement.

## Reference Contours

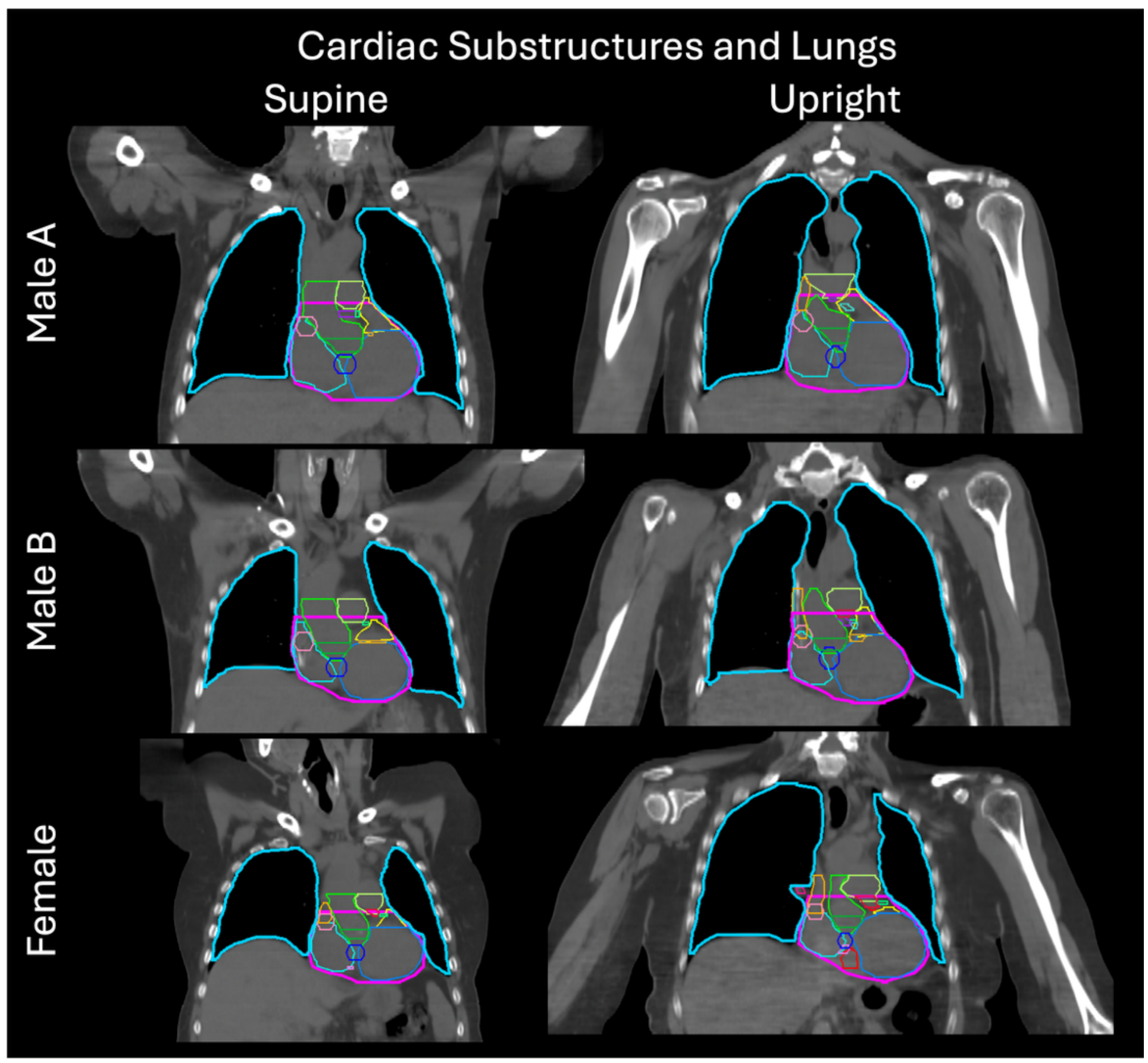

Figure 2: Cardiac substructures included in the assessment overlaid on three arbitrary patient average 4D CT images showing relative differences between supine and upright patient positioning.

Average CT images were manually labeled with 20 cardiac substructures including the whole heart (WH); chambers, left/right atria/ventricles (LA, RA, LV, RV); great vessels (GVs), ascending aorta (AA), superior/inferior vena cava (SVC, IVC), and pulmonary artery/veins (PA,

PVs); CAs, left main (LMCA), left anterior descending (LADA), right (RCA), and left circumflex (LCx) arteries; valves, aortic (V-AV), pulmonic (V-PV), tricuspid (V-TV), and mitral (V-MV) valves; and conduction nodes, sinoatrial (N-SA) and atrioventricular (N-AV) nodes following a consensus of published guidelines[18–22]. Expert consultation was performed as needed with a radiologist with cardiovascular subspecialty and 10+ years of experience. In addition to the CS, the lungs were manually contoured[23] to enable volumetric comparisons between positions. Representative paired supine and upright images are shown in Figure 2 highlighting CS overlaid on patient images.

To further evaluate relative treatment geometries, the carina was utilized as a clinically relevant thoracic landmark for evaluating posture-dependent changes in cardiac position near central and ultra-central targets[24,25]. The carina was represented by a manually contoured volumetric surrogate region on axial CT images to improve landmark consistency. The contour began at the slice immediately superior to the tracheal bifurcation, where the distal tracheal lumen first began to widen toward the mainstem bronchi, and continued inferiorly through the bifurcation until the right and left mainstem bronchi were visualized as two discrete luminal structures[26]. Using this structure, the centroid was extracted to represent a consistent image-based reference point for comparison[27].

## Analysis

A previously described, in-house DL segmentation model[22], trained on supine-positioned images, was applied to each average CT image from the eight paired upright and supine cases to generate automatic contours for the 20 CS. Model performance was compared between upright and supine images to evaluate differences in auto-segmentation performance under posture-related anatomical changes. Predicted contours were compared against the manually defined reference contours using the Dice similarity coefficient (DSC) for contour overlap and 95% Hausdorff distance (HD95) for spatial agreement following published guidelines[28]. For each structure and performance metric, upright and supine model performance was compared using a two-tailed Wilcoxon signed-rank test with $p<0.05$ considered statistically significant. Predictions from both positions were qualitatively visualized with the reference contours overlaid on the underlying anatomy.

To enable posture-dependent centroid comparisons in a common coordinate frame, each upright average CT was rigidly registered to the corresponding supine image using a box-based assisted alignment algorithm in MIM Maestro (version 7.4.2, MIM Software Inc., Cleveland, OH) and manually adjusted, ensuring alignment of the central vertebral bodies as anatomical anchors. Lung and WH volumes were compared between upright and supine average CTs for each patient. For each cardiac substructure, centroid displacement from supine to upright was quantified in the registered coordinate frame for directional displacement comparisons along each major axis. Cardiac substructure centroid positions were also examined relative to the carina centroid. For each posture-dependent comparison, two-tailed one-sample Wilcoxon signed rank tests were used to determine whether the paired difference differed significantly from zero, with $p<0.05$ considered statistically significant.

End-inhalation and end-exhalation phase images were analyzed on the subset of three patients for the WH, chambers, AA, and CAs, in both positions to explore posture-dependent differences in CS stability during respiration when upright. For each patient and position, respiratory-induced motion was quantified as the centroid displacement of each included structure between end-exhalation and end-inhalation and was compared descriptively between upright and supine positions. Due to the limited sample size and exploratory nature of the work, no statistical testing was performed for the respiratory-motion analysis. Results were summarized descriptively as an exploratory assessment of posture-dependent respiratory motion.

## Results

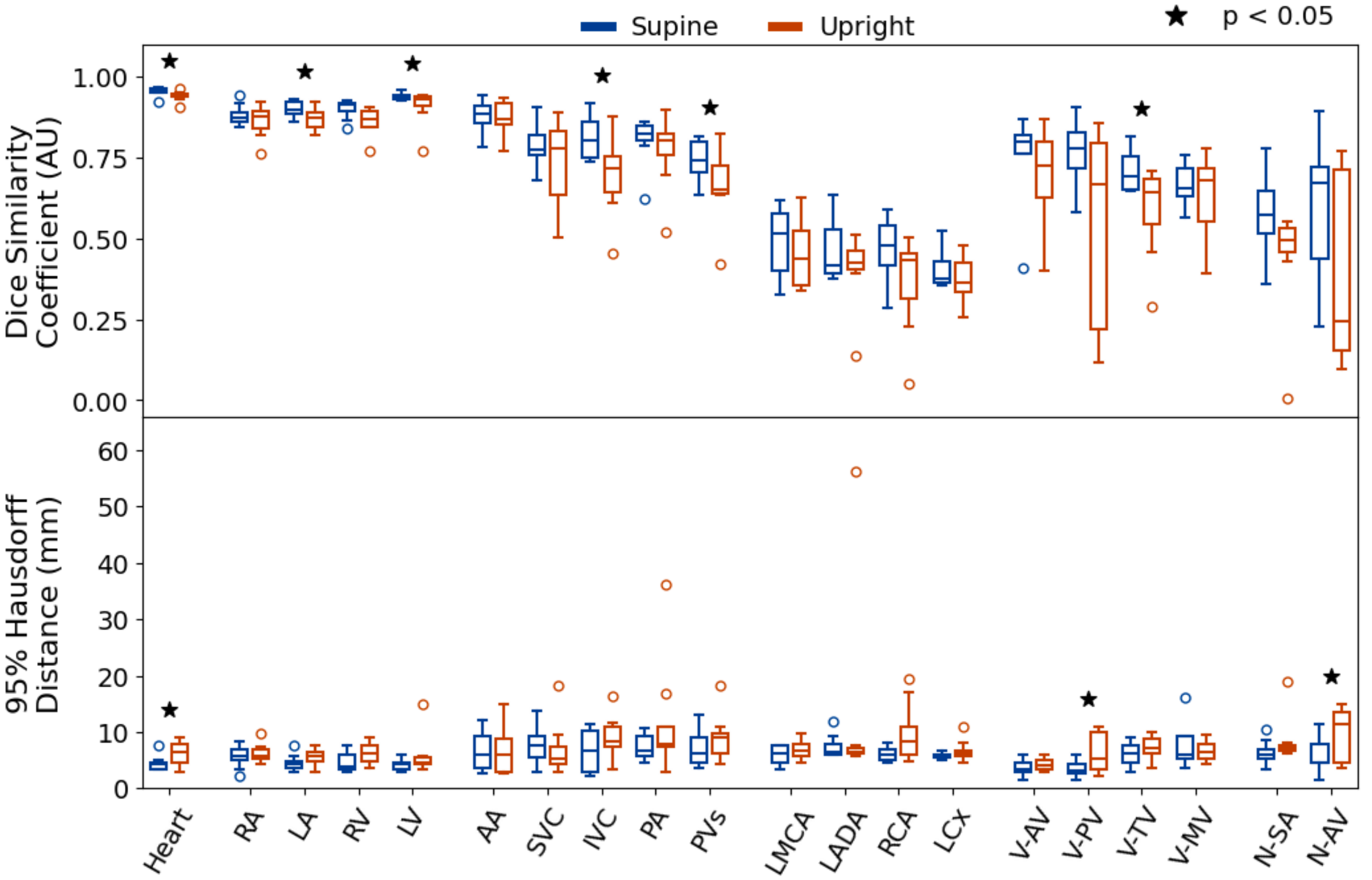

Figure 3: Quantitative performance of a deep learning model, trained on a supine dataset for 20 cardiac substructures applied to both supine and upright average 4DCT images, evaluated using the Dice similarity coefficient and 95% Hausdorff distance. Abbreviations: Arbitrary Units, AU; remaining abbreviations are defined in text.

Quantitative performance of the pre-trained DL model on supine and upright CT images is shown in Figure 3. Across all 20 CS for 8 matched thoracic cancer patients, average DSC was 0.72±0.19 (heart, 0.95±0.01; chambers, 0.91±0.03; GVs, 0.80±0.08; CAs, 0.46±0.09; valves, 0.72±0.11; nodes, 0.59±0.17) for supine images and 0.65±0.24 (heart, 0.94±0.02; chambers, 0.88±0.05; GVs, 0.75±0.13; CAs, 0.40±0.11; valves, 0.61±0.20; nodes, 0.41±0.23) for upright images. Average HD95 across all 20 structures was 5.8±2.6mm (heart, 4.5±1.3mm; chambers, 4.7±1.6mm; GVs, 7.1±3.3mm; CAs, 6.3±1.6mm; valves, 5.2±2.9mm; nodes, 6.1±2.6mm) for supine images and 7.6±5.6mm (heart, 6.3±2.1mm; chambers, 6.0±2.3mm; GVs, 8.8±6.0mm; CAs, 9.0±9.1mm; valves, 6.1±2.5mm; nodes, 9.1±4.3mm) for upright images. Segmentation performance was significantly different ($p<0.05$) for supine images than upright images, corresponding to a DSC decrease of 0.07 and HD95 increase of 1.8mm when applied to upright images. Despite this performance reduction, the model successfully generated complete predictions of all CS in all input images.

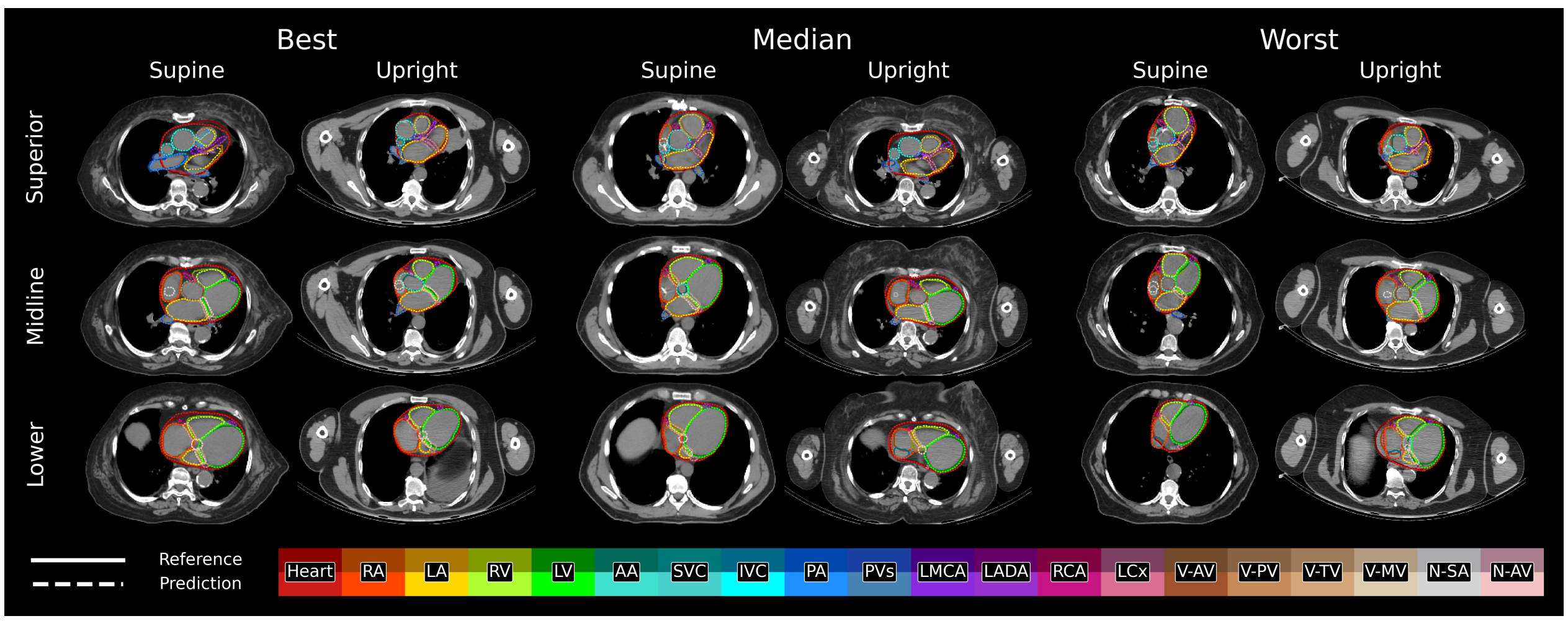

Figure 4: Qualitative segmentation results for supine and upright averaged 4D CT-SIM images generated by a deep learning model pre-trained on a supine dataset for 20 cardiac substructures. Best, median, and worst cases were selected separately for each position based on substructure-average Dice similarity coefficient.

Qualitative axial view comparisons are shown in Figure 4. Best, median, and worst cases were selected separately for the supine and upright datasets based on average DSC across all substructures. Predicted contours generally aligned well with the reference contours for larger structures, although visible discrepancies were present across both positions, particularly near the superior and inferior boundaries of individual structures and where one structure transitions into an adjacent one, such as the RA-SVC interface. In the median and worst cases, upright predictions demonstrated greater contour variability than supine, including underprediction along cardiac borders and misclassification within image regions with homogeneous signal intensity and a lack of distinct borders. Nevertheless, predictions in both positions generally produced complete, anatomically plausible contours relative to the underlying image features, suggesting that the model may provide a useful starting point for contour review and refinement in treatment-planning workflows for upright RT.

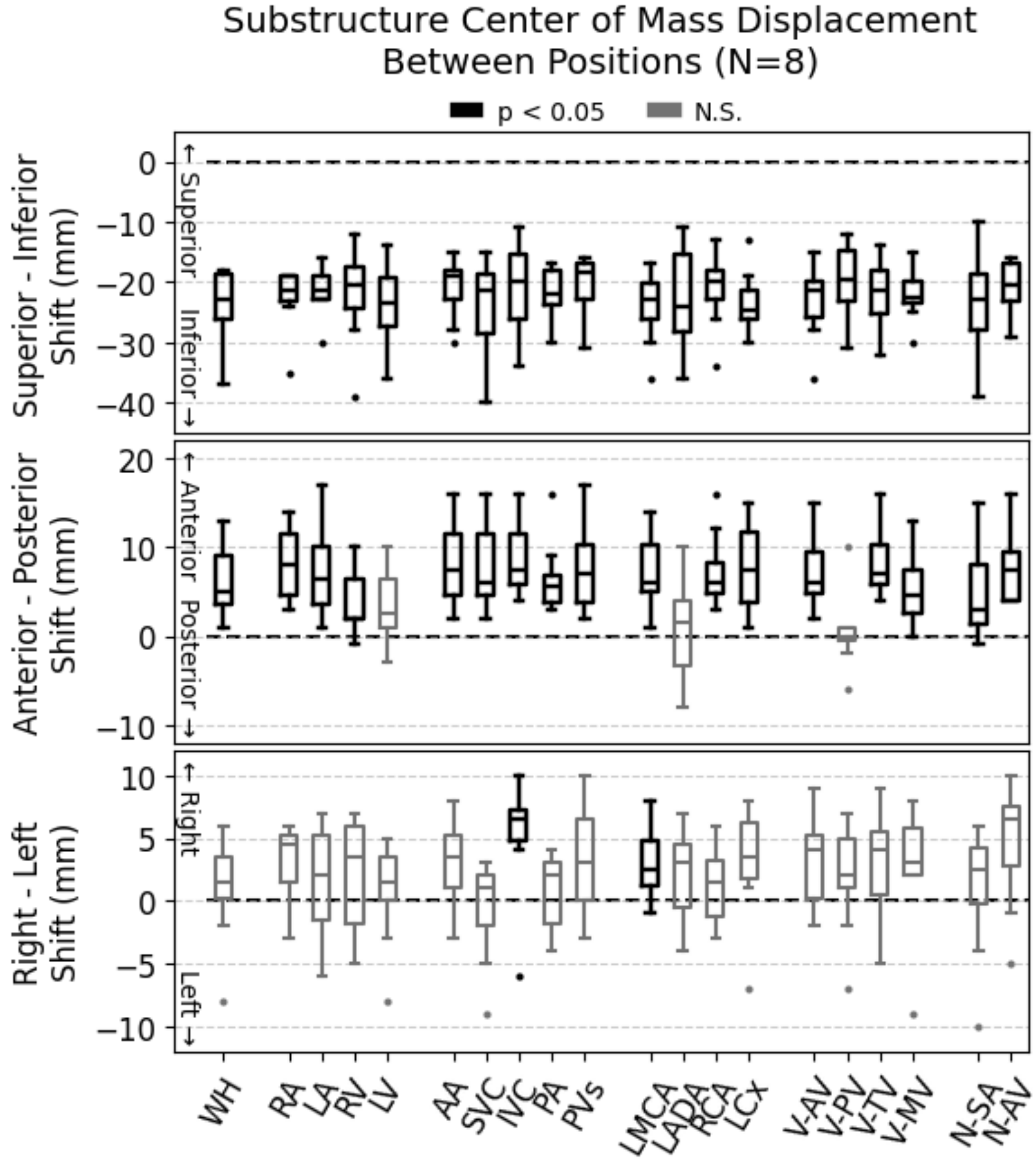

Figure 5: Cardiac substructure centroid displacement measuring structure shift from supine to upright positions along each dimension for eight patients. Abbreviations: non-significant, N.S.; remaining abbreviations are defined in text.

Across the eight evaluated patients, lung volume increased significantly ($p<0.05$) when upright, with a median increase of 20.6% (range, -8.9% – 42.8% increase) relative to supine imaging. Only one patient demonstrated a decrease in lung volume when positioned in the upright orientation. No significant differences in WH volume were observed in the different patient positions (median decrease of 4.3% (range, -4.0% – 14.8% reduction)relative to supine patient position. CS centroid displacement between registered supine and upright images are shown in Figure 5. When upright, all CS demonstrated significant inferior displacement of at least 10mm with a median WH inferior shift of 23mm (range, 18mm – 37mm). Most structures also shifted anteriorly, with statistically significant ($p< 0.05$) anterior displacement differences observed for all structures except the LV, LADA, and V-PV. The median WH anterior shift of 5.0mm (range, 1.0mm – 13.0mm) whereas the median displacements were between 4mm and 25.5mm anterior. No significant differences in left-right displacements were observed, with median structures displacements between 3mm right and 2mm left.

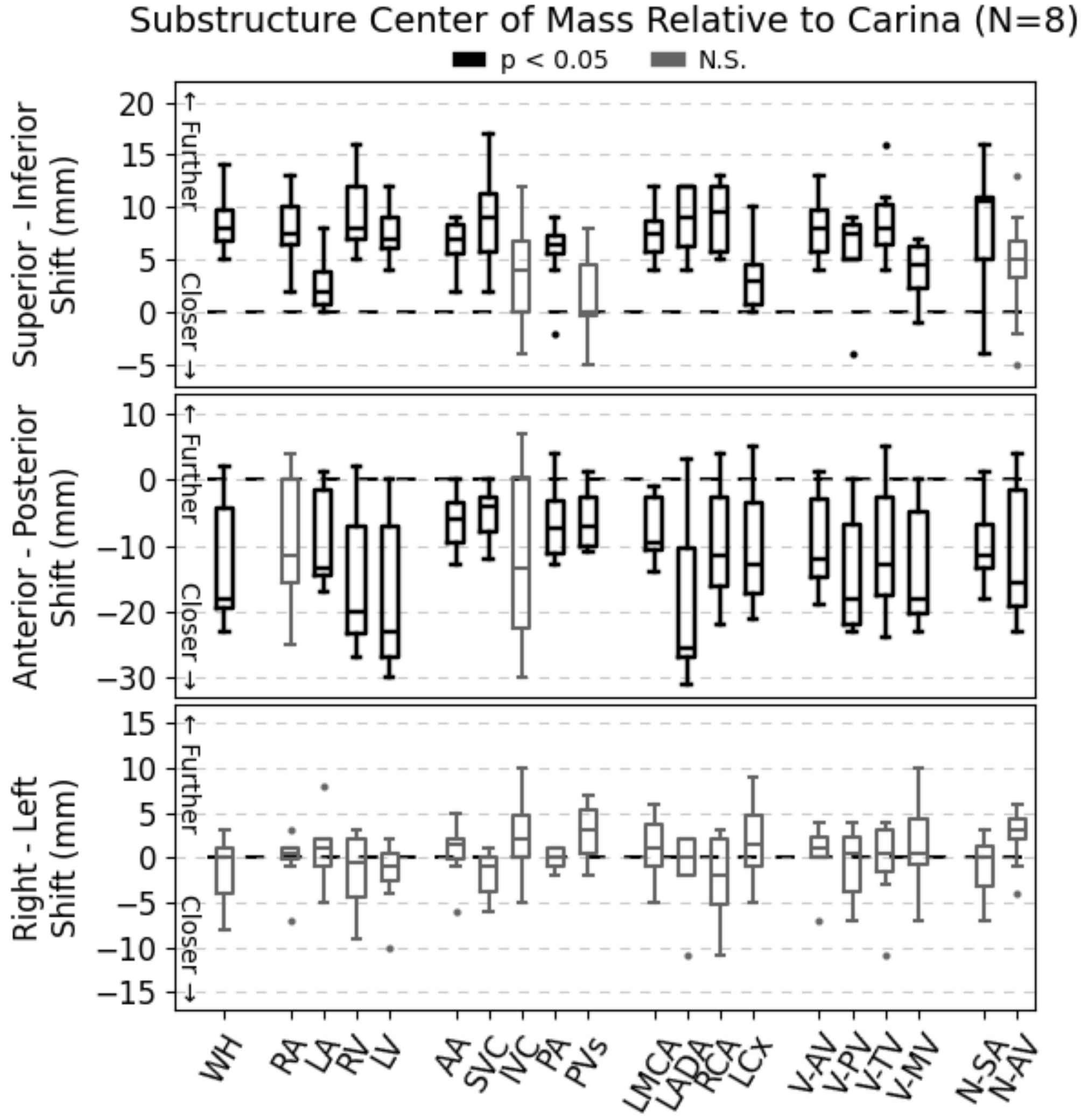

Figure 6: Cardiac substructure centroid displacement relative to the carina between supine and upright imaging. Abbreviations: Non-Significant, N.S.; remaining abbreviations are defined in text.

CS centroid positions relative to the carina centroid are shown in Figure 6. When upright, most CS were positioned significantly farther inferior relative to the carina, except for the IVC and PVs. The median WH shift was 8mm (range, 5mm – 14mm) with a range of median CS shifts between 0.0mm and 10.5mm further inferior with respect to the carina. In the anterior-posterior direction, CS shifted closer to the carina when upright, with significant displacements observed for most structures. The median WH shift was 18mm (range, -2mm – 23mm) with a range of CS shifts between 4mm and 25.5mm closer in-line anterior-posterior with the carina. Left-right displacements relative to the carina were smaller, non-significant, and distributed around zero with a range of CS shifts between 2mm closer and 3mm further away from the carina.

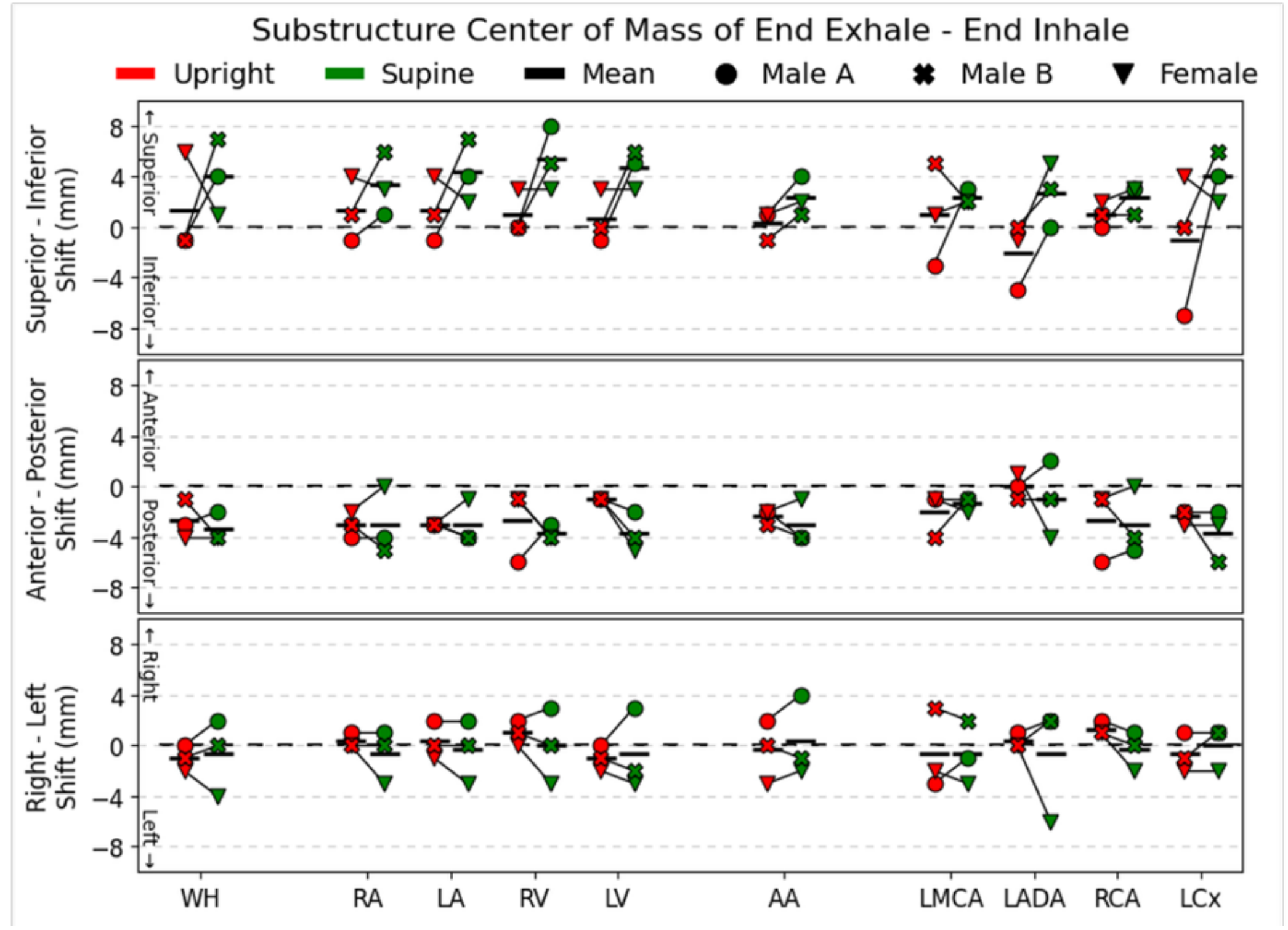

Figure 7: Cardiac substructure centroid displacement between end-inhalation and end-exhalation for a limited subset of three patients, comparing respiratory-driven cardiac motion between supine and upright positioning.

CS centroid displacement between end-inhalation and end-exhalation is shown in Figure 7 for the three patients with respiratory phase images in both supine and upright positions. Respiratory-driven displacement was primarily observed in the superior-inferior direction, while anterior-posterior and left-right displacements were minimal across patients with average displacements less than 3mm for all structures. The magnitude and direction of posture-related differences in respiratory motion varied across patients. For male A and male B, upright positioning was associated with reduced whole-heart superior-inferior displacement between respiratory phases by 3mm and 6mm, respectively. In contrast, the third patient demonstrated a 5mm increase in whole-heart superior-inferior displacement when upright. These patient-specific differences appeared to correspond with changes in lung volume between positions, as the two patients with reduced upright displacement demonstrated 24.5% and 24.0% increase in lung volume, whereas the patient with increased upright displacement demonstrated a 5.3% increase.

## Discussion

This study evaluated posture-related changes in CS geometry between supine and upright positioning for thoracic RT patients. Across the cohort, a DL model, pre-trained on supine based images[22] was successfully able to predict all CS on all upright images, with a slightly reduced performance of 0.07 DSC (average supine DSC, 0.72±0.19; average upright DSC, 0.65±0.24). When upright, the heart demonstrated a dominant inferior displacement of ~2cm and a reduced anterior displacement of ~0.5cm relative to the registered supine image. Relative to the carina, the heart was positioned 0.7cm farther inferior but 1.3cm closer in the anterior-posterior direction when upright. In the limited subset of patients with respiratory phase images available, reduced respiratory-driven cardiac displacement was observed in two patients who demonstrated larger upright increases in lung volume extent (~25%). However, this pattern was not observed in the patient with more limited increase (~5%). Together, these findings suggest that upright positioning may produce more favorable cardiac geometry for some thoracic RT patients by increasing separation between central and ultra-central target-adjacent anatomy[24,25] and radiosensitive cardiac substructures, while remaining applicable to existing, validated supine-based DL models.

A previously validated, DL segmentation model trained on supine CT images was applied to upright CT images to evaluate generalizability across patient position[22]. When applied to supine images from this cohort, the model achieved an average DSC of 0.72 across all 20 CS, which is comparable to previously reported CS segmentation studies using supine CT images[16,22]. When applied to upright images, a statistically significant reduction in performance was observed with average DSC decreasing by ~0.07. However, the corresponding spatial differences in CS were modest with average HD95 increasing by ~2mm from supine to upright images. These results indicate that upright imaging may introduce slight changes in segmentation performance for supine-trained segmentation models. Yet, the model still produced complete and anatomically plausible contours for all evaluated structures. Importantly, the observed performance difference cannot be attributed solely to patient posture because the upright and supine images were acquired using different CT scanners and imaging systems, which may also affect image contrast, noise, texture, resolution, and overall image quality[17]. Nevertheless, the ability of the model to generate complete predictions despite upright-related geometric changes, including inferior cardiac displacements, potential cardiac rotation, patient arm position, and the 20° chair tilt, suggests that supine-trained models may remain applicable to upright imaging and provide a useful starting point for efficient contour review and refinement in upright RT workflows.

The posture-related anatomical changes observed in this study were generally consistent with prior upright imaging literature. Across the eight thoracic RT patients, lung volume increased by a median of approximately 21% when upright, with only one patient demonstrating a decrease in lung volume. Yamada *et al.*[9] reported upright lung volume increases of approximately 5 to 15% in 100 asymptomatic volunteers, while Yang *et al.*[6] reported an average upright lung volume increase of approximately 27% in a radiotherapy simulation study of five healthy volunteers. Although the patients in this study were being treated for advanced thoracic cancer, which may influence posture-related lung expansion, the observed lung volume changes were within the range of prior reports. Similar agreement was observed for cardiac displacement. Marano *et al.*[8] reported that the inferior aspect of the heart shifted approximately 4mm farther away from the superior extent of the sternum when upright in 15 RT patients, while Kondaveeti *et al.*[29] reported an approximately 16mm inferior shift of the LV apex relative to the nearest intercostal space in 10 asymptomatic volunteers. Although these studies used different anatomical landmarks, this work similarly demonstrated a dominant inferior WH centroid shift of ~20mm between registered supine and upright images, and an ~10mm inferior shift relative to the carina. Finally, Norimatsu *et al.*[30] reported that the anatomical cardiac axis moved dorsally and caudally from supine to upright positioning. While this work did not directly evaluate cardiac axis orientation, the observed inferior and posterior shifts relative to the carina may reflect similar posture-dependent changes in cardiac orientation.

Although limited to three patients, the respiratory phase analysis suggests that the effect of upright positioning on respiratory-driven cardiac displacement may depend on patient-specific breathing mechanics. The two patients with reduced upright cardiac displacement also demonstrated larger upright lung volume increases, whereas the patient with increased upright displacement demonstrated only a modest lung volume increase. This pattern is consistent with prior upright thoracic imaging studies suggesting that on average, respiratory motion may be reduced when positioned upright[5,6]. These findings were also broadly consistent with respiratory motion management literature suggesting that internal anatomy stability during thoracic RT can be influenced by diaphragm motion and breathing pattern variability[3,4]. Therefore, upright positioning may offer a motion-management advantage for some patients. Future studies with large respiratory-resolved cohorts should evaluate how variable lung volume expansion may correlate with patients experiencing reduced respiratory-driven cardiac displacement when upright.

An important limitation of this study is the small sample size. Paired upright and supine CT datasets remain limited because upright RT is still emerging, and such datasets are not yet widely available. The patient population evaluated had advanced diseases that required individualized setup and immobilization for comfort and patient support structures that may not reflect broader thoracic RT populations. In addition, upright imaging was performed using the Brilliance Big Bore CT, configured for upright imaging by P-Cure, with a fixed 20° chair tilt. Therefore, the measured deformations may not exactly represent those expected for patients positioned at ±15° upright orientation available on other upright RT systems. However, the magnitude and direction of the observed changes were consistent with prior paired upright-supine imaging studies, including studies using 0° chair tilt[6,9]. The dominant anatomical effect is expected to arise from the transition from supine positioning to upright due to sources such as gravity-dependent shifts, organ-to-organ interactions, and organ deformation. Therefore, although the exact deformation field may vary across different upright systems with different chair tilt angles available, the measured inferior cardiac displacement and associated implications for cardiac sparing remain relevant to upright RT workflow development.

Another limitation of this work is that it focused primarily on lung and mediastinal cancer patients, and additional evaluation is needed across other disease sites where cardiac sparing is clinically relevant, including breast, thymoma, and esophageal cancers. This is especially important as proton and particle therapy have been investigated as cardiac-sparing strategies across several thoracic and thoracic-adjacent sites. In breast cancer, the RADCOMP trial[31] is prospectively comparing proton and photon therapy with cardiac outcomes as a major endpoint, and proton therapy has demonstrated low whole-heart and CS dose in breast cancer patients[32]. In esophageal cancer, proton therapy has been shown to reduce whole-heart and CS dose compared with intensity-modulated RT (IMRT)[33], with broader evidence suggesting reduced heart, LV, and LADA dose compared with photon RT[34]. Similarly, in thymoma or anterior mediastinal tumors, proton therapy has demonstrated reduced heart dose compared with IMRT[35], with recent work also showing improved CS sparing using intensity-modulated PT[36]. Although early upright particle therapy work has suggested potential dosimetric benefits for lung cancer[10], CS-sparing across upright disease-site applications remains insufficiently studied.

Future work will include dosimetric evaluation of these patients to better contextualize the dosimetric impact of the measured geometric shifts for RT planning. In a recent study, Martire *et al.*[10] demonstrated carbon-ion treatment planning in a similar cohort of thoracic patients and found that dosimetric comparisons between positions were highly dependent on patient-specific anatomy. In the present work, the availability of CS contours on the average CT images used for RT planning creates an opportunity for direct investigation of advanced cardiac sparing opportunities with upright RT including particle and photon therapy. Such analysis could better determine whether the observed upright geometric changes translate into clinically meaningful improvements in treatment geometry and cardiac dose reduction in support of future prospective study design.

## Conclusion

This study demonstrated that upright positioning produces measurable posture-related changes in CS geometry for thoracic RT patients, including dominant inferior WH displacement, altered positioning relative to the carina, and increased lung volume in most patients. A supine-trained CS segmentation model showed slightly reduced performance on upright CT images but still generated anatomically plausible contours suitable for clinical review and refinement. Although limited by small sample size, scanner differences, and heterogeneous patient setup, these findings support further investigation of upright RT for improving thoracic treatment geometry and cardiac sparing.